\documentclass[runningheads]{svmult}

\usepackage{makeidx}   
\usepackage{graphicx}  
\usepackage{subeqnar}  
\usepackage{multicol}  
\usepackage{physprbb}  
\makeindex             

\catcode`\@=11
\def\gsim{\ifmmode{\mathrel{\mathpalette\@versim>}}
    \else{$\mathrel{\mathpalette\@versim>}$}\fi}
\def\lsim{\ifmmode{\mathrel{\mathpalette\@versim<}}
    \else{$\mathrel{\mathpalette\@versim<}$}\fi}
\def\@versim#1#2{\lower 2.9truept \vbox{\baselineskip 0pt \lineskip 
    0.5truept \ialign{$\m@th#1\hfil##\hfil$\crcr#2\crcr\sim\crcr}}}
\catcode`\@=12
\def\Mbh{M_{\rm BH}}
\def\Mbht{\Mbh^{\rm T}}
\def\Nbh{N_{\rm BH}^>}
\def\Lq{L_{\rm Q}}
\def\Lqt{\Lq^{\rm T}}

\def\Etq{E^{\rm T}_{\rm Q}}
\def\Nq{N_{\rm Q}^>}
\def\fq{f_{\rm Q}}
\begin{document}
\title*{What does the local black hole mass distribution tell us 
about the evolution of the quasar luminosity function?}
\titlerunning{An estimate for QSO duty cycle}
\author{Luca Ciotti\inst{1,2,3}
\and Zoltan Haiman\inst{3}
\and Jeremiah P. Ostriker\inst{4,3}}
\authorrunning{Luca Ciotti et al.}
%
\institute{Osservatorio Astronomico di Bologna, via Ranzani 1, 
           40127 Bologna, Italy
\and Scuola Normale Superiore di Pisa, Piazza dei Cavalieri 7, 56126 Pisa, 
     Italy
\and Princeton University Observatory, Peyton Hall, 08544 NJ, USA
\and Institute of Astronomy, Cambridge University, Madingley Road, CB3 0HA, 
     Cambridge, UK}
\maketitle              

\begin{abstract}
We present a robust method to derive the duty cycle of QSO activity
based on the empirical QSO luminosity function and on the present-day
linear relation between the masses of supermassive black holes and
those of their spheroidal host stellar systems.  It is found that the
duty cycle is substantially less than unity, with characteristic
values in the range $3-6\times 10^{-3}$. Finally, we tested the
expectation that the QSO luminosity evolution and the star formation
history should be roughly parallel, as a consequence of the
above--mentioned relation between BH and galaxy masses.
\end{abstract}

\section{Introduction}
The discovery of remarkable correlations between the masses of
supermassive BHs hosted at the centers of galaxies and the global
properties of the parent galaxies themselves~\cite{8,9,12} leads to a
natural link between the cosmological evolution of QSOs and the
formation history of galaxies~\cite{2}. The investigation of such
interesting correlations looks promising not only to better understand
how and when galaxies formed, but also to obtain information on the
QSO population itself~\cite{4}. Here we focus on two specific points
raised by the general remarks above: 1) The use of the ``Magorrian
relation'' to determine the QSO duty cycle at redshift $z=0$; 2) The
expected relation between the cosmological evolution of the total
luminosity emitted by star--forming galaxies and that of the total
luminosity emitted by QSOs.  As we will see, an interesting
consequence of this last point is the possible existence of a physical
process limiting gas accretion onto BHs at high redshifts.  The {\it
observational} inputs of our analysis are the Magorrian
relation~\cite{12}, the galaxy mass--to--light ratio (from the
Fundamental Plane)~\cite{5,6}, the present-day luminosity function of
spheroids~\cite{14}, the present-day and the integrated QSO
cosmological (light) evolution~\cite{13}, and finally the star
formation history~\cite{11}.  A possible alternative to the use of the
mass-to-light ratio is the use of the Faber-Jackson~\cite{7} relation
coupled with the so-called $\Mbh-\sigma$ relation~\cite{8,9}.  The
technical details will be given elsewhere~\cite{4}.
\section{Results}
We start our analysis by deriving the QSO's mean accretion efficiency
associated with the BH's growth (see, e.g.,~\cite{15}) as $\epsilon
\equiv \Etq /\Mbht c^2\simeq 0.06$, where $\Mbht$ is the present-day
total mass of BHs at the center of stellar spheroids, and $\Etq$ is
the total energy emitted by all QSOs over the entire life of the
Universe. We then derive the QSO (mean) duty cycle $\fq$  in two ways:
\begin{equation}
<\fq>_x\equiv {\epsilon c^2 \Mbh (x)\over \int_0^{\infty} \Lq (x,t)dt},
\quad
<\fq>_x\equiv {\Nbh[\Mbh (x)] \over\Nq[\Lq(x,0)]},
\end{equation}
where $\Lq(x,t)$ is such that QSOs brighter than $\Lq(x,t)$ (whose
number is $\Nq[\Lq(x,t)]$) emit the fraction $x$ of $\Lqt(t)$, and
$\Mbh(x)$ is such that all BHs heavier than $\Mbh(x)$ (whose number is
$\Nbh[\Mbh(x)]$) sum up to $x\Mbht$ (at z=0). We found that the two
ways give consistently $<\fq>_{0.1}\simeq 0.003$ and
$<\fq>_{0.9}\simeq 0.006$, in good agreement with theoretical
results~\cite{1}. Our result is similar in spirit but different in
detail from that derived in~\cite{10}, who obtained a {\it QSO
lifetime} of $\simeq 10^7$ yr at a Hubble epoch of $\simeq 10^9$ yr
or, in our terms, $\fq\simeq 10^{-2}$.  Finally, an interesting
expectation based on the Magorrian relation would be a parallel
evolution of the QSO accretion and star formation histories.  As is
well known, at $z\lsim 2$ the QSO total luminosity and the UV
luminosity associated to the star formation history are indeed roughly
parallel but at $z\gsim 2$ they have clearly divergent slopes. A fit
shows that the QSO luminosity evolution is reasonably well fitted
under the assumptions that BHs are accreting at {\it one tenth} of the
Eddington luminosity, as computed under the standard assumption of
pure electron scattering. A global picture could then be that at low
redshift the BH accretion (and star formation) are limited by the
available amount of gas, while at high redshift some extra source of
opacity (as for example bremsstrahlung opacity~\cite{3}), able to
reduce the Eddington luminosity of one order of magnitude (with
respect to the pure electron scattering case), is at work~\cite{4}.

\end{document}